\documentclass[8.5pt,twoside]{article}

\usepackage[super,sort&compress,comma]{natbib} 
\usepackage{times,mathptm}
\usepackage{sectsty}
\usepackage{balance} 
\usepackage[text={11.3cm,20.4cm},centering]{geometry} 
\usepackage{color}

\usepackage{graphicx} 
\usepackage{lastpage}
\usepackage[format=plain,justification=raggedright,singlelinecheck=false,font=small,labelfont=bf,labelsep=space]{caption} 
\usepackage{fancyhdr}
\begin{document}

\thispagestyle{plain}
\fancypagestyle{plain}{
\fancyhead[L]{\includegraphics[height=8pt]{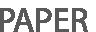}}
\fancyhead[C]{\hspace{-1cm}\includegraphics[height=15pt]{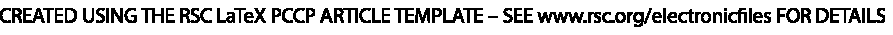}}
\fancyhead[R]{\includegraphics[height=10pt]{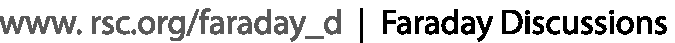}\vspace{-0.2cm}}
\renewcommand{\headrulewidth}{1pt}}
\renewcommand{\thefootnote}{\fnsymbol{footnote}}
\renewcommand\footnoterule{\vspace*{1pt}%
\hrule width 11.3cm height 0.4pt \vspace*{5pt}} 
\setcounter{secnumdepth}{5}

\makeatletter 
\renewcommand{\fnum@figure}{\textbf{Fig.~\thefigure~~}}
\def\subsubsection{\@startsection{subsubsection}{3}{10pt}{-1.25ex plus -1ex minus -.1ex}{0ex plus 0ex}{\normalsize\bf}} 
\def\paragraph{\@startsection{paragraph}{4}{10pt}{-1.25ex plus -1ex minus -.1ex}{0ex plus 0ex}{\normalsize\textit}} 
\renewcommand\@biblabel[1]{#1}            
\renewcommand\@makefntext[1]%
{\noindent\makebox[0pt][r]{\@thefnmark\,}#1}
\makeatother 
\sectionfont{\large}
\subsectionfont{\normalsize} 

\fancyfoot{}
\fancyfoot[LO,RE]{\vspace{-7pt}\includegraphics[height=8pt]{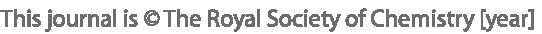}}
\fancyfoot[CO]{\vspace{-7pt}\hspace{5.9cm}\includegraphics[height=7pt]{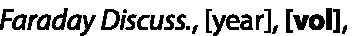}}
\fancyfoot[CE]{\vspace{-6.6pt}\hspace{-7.2cm}\includegraphics[height=7pt]{RF}}
\fancyfoot[RO]{\scriptsize{\sffamily{1--\pageref{LastPage} ~\textbar  \hspace{2pt}\thepage}}}
\fancyfoot[LE]{\scriptsize{\sffamily{\thepage~\textbar\hspace{3.3cm} 1--\pageref{LastPage}}}}
\fancyhead{}
\renewcommand{\headrulewidth}{1pt} 
\renewcommand{\footrulewidth}{1pt}
\setlength{\arrayrulewidth}{1pt}
\setlength{\columnsep}{6.5mm}
\setlength\bibsep{1pt}

\noindent\LARGE{\textbf{The cycling of carbon into and out of dust}} 
\vspace{0.6cm}

\noindent\large{\textbf{Anthony P. Jones,$^{\ast}$$^{a}$, Nathalie Ysard$^{a}$, Melanie K\"ohler$^{a}$, \\ Lapo Fanciullo$^{a}$, Marco Bocchio$^{a}$, Elisabetta Micelotta$^{a}$, \\ Laurent Verstraete$^{a}$ and Vincent Guillet$^{a}$}\vspace{0.5cm}

\noindent\textit{\small{\textbf{Received Xth XXXXXXXXXX 20XX, Accepted Xth XXXXXXXXX 20XX\newline
First published on the web Xth XXXXXXXXXX 200X}}}

\noindent \textbf{\small{DOI: 10.1039/c000000x}}
\vspace{0.6cm}

\noindent \normalsize{Observational evidence seems to indicate that the depletion of interstellar carbon into dust shows rather wide variations and that carbon undergoes rather rapid re-cycling in the interstellar medium (ISM). Small hydrocarbon grains are processed in photo-dissociation regions by UV photons and by ion and electron collisions in interstellar shock waves and by cosmic rays. A significant fraction of hydrocarbon dust must therefore be re-formed by accretion in the dense, molecular ISM. A new dust model [Jones \textit{et al., A\&A}, 2013, \textbf{558}, A62]} shows that variations in the dust observables, in the diffuse interstellar medium ($n_{\rm H} \leq 10^3$\,cm$^{-3}$), can be explained by systematic and environmentally-driven changes in the small hydrocarbon grain population. Here we explore the consequences of gas-phase carbon accretion onto the surfaces of grains in the transition regions between the diffuse ISM and molecular clouds [\textit{e.g.}, Jones, \textit{A\&A}, 2013, \textbf{555}, A39]. We find that significant carbonaceous dust re-processing and/or mantle accretion can occur in the outer regions of molecular clouds and that this dust will have significantly different optical properties from the dust in the adjacent diffuse ISM. We conclude that the (re-)processing and cycling of carbon into and out of dust is perhaps the key to advancing our understanding of dust evolution in the ISM. 
\vspace{0.5cm}

\footnotetext{\textit{$^{a}$~Institut dÕAstrophysique Spatiale, CNRS/Universit\'e Paris Sud, Orsay, 91405, France. Fax: 33 1 6985 8675; Tel: 33 1 6985 8647; E-mail: Anthony.Jones@ias.u-psud.fr}}

\section{Introduction}

Many lines of observational evidence and much modelling work indicate that interstellar carbon grains and, in particular, carbonaceous nano-particles are rather `volatile' dust species that seemingly undergo rapid processing in the interstellar medium (ISM).\cite{2008A&A...492..127S,2010A&A...510A..36M,2010A&A...510A..37M,2011A&A...526A..52M,2011A&A...530A..44J,2012ApJ...760...36P,2012A&A...542A..69P,2012A&A...545A.124B}  
Recent work strongly suggests that these grains most likely consist of hydrogenated amorphous carbons, a-C(:H), which encompass H-poor, aromatic-rich a-C through to H-rich, aliphatic-rich a-C:H materials.\cite{2012A&A...540A...1J,2012A&A...540A...2J,2012A&A...542A..98J,2012A&A...545C...2J,2012A&A...545C...3J,2013A&A...555A..39J, 2013A&A...558A..62J} Such hydrocarbon nano-particles are processed in the intense radiation fields of photo-dissociation regions (PDRs) by UV photon-induced destruction\cite{2012A&A...542A..69P, 2013A&A...558A..62J} and by ion and electron collisions.  
This processing can occur in interstellar shock waves, in a hot gas ($T \geq 10^6$\,K) and by cosmic ray interactions.\cite{2008A&A...492..127S,2010A&A...510A..36M,2010A&A...510A..37M,2011A&A...526A..52M,2012A&A...545A.124B,2013A&A...556A...6B}
Given that a significant fraction of carbon is in dust in the diffuse ISM, rapid hydrocarbon dust destruction implies that it must be re-formed by accretion in the dense, molecular ISM.\cite{2011A&A...530A..44J}  

Here we adopt a new dust model,\cite{2013A&A...558A..62J} which can explain many of the variations in the dust observables (UV-NIR extinction and IR-mm emission) in the diffuse interstellar medium ($n_{\rm H} \leq 10^3$\,cm$^{-3}$) in terms of systematic and environmentally-driven changes in the small hydro-carbon grain population. Further, the photo-processing of newly-exposed a-C:H nano-particles, in the shells around evolved stars, could provide a viable mechanism for fullerene formation.\cite{2012ApJ...757...41B,2012ApJ...761...35M}  We study the accretion of gas-phase carbon and briefly consider the formation of aggregates by the coagulation of small a-C(:H) grains onto the surfaces of large grains in the transition regions between the diffuse ISM and molecular clouds.\cite{2003A&A...398..551S,2012A&A...548A..61K,2011A&A...528A..96K,2013A&A...559A.133Y} We conclude that the processing and cycling of carbon into and out of dust appears to be the key to advancing our understanding of dust evolution in the ISM.

\section{The evolution of carbonaceous dust in the ISM}

In the laboratory a-C(:H) solids are known to darken upon UV photon irradiation, thermal annealing and ion irradiation.\cite{1985OpEffinAS.120..258I,1984JAP....55..764S,2011A&A...529A.146G} In a-C(:H) materials this process, known as photo-darkening\footnote{Photo-darkening refers to an increase in the {\em dark}, graphite-like or aromatic content. Hence, the term photo-darkening is used to describe the aromatisation of a-C(:H) materials.}, leads to a decrease in the band gap or optical gap energy, $E_{\rm g}$. The band gap for a-C(:H) materials ($E_{\rm g} = -0.1$ to 2.7\,eV)\footnote{Here we adopt the Tauc gap for the optical band gap, $E_{\rm g}$, {\it i.e.}, the energy axis intercept in a plot of  $(E \, \alpha)^{0.5} \ vs. \ E$, where $\alpha = 4\pi k/\lambda$ is the absorption coefficient.\cite{2012A&A...540A...2J}} is directly related to $X_{\rm H}$, the H atom fraction ($E_{\rm g} = 4.3 X_{\rm H}$) and also to the ratio, $R$, of the $sp^3$ and $sp^2$ C atomic fractions, $X_{sp^3}$ and $X_{sp^2}$, respectively, {\it i.e.},\cite{2012A&A...540A...1J,2012A&A...540A...2J} 
\begin{equation}
R = \frac{X_{sp^3}}{X_{sp^2}} \approx \frac{(8X_{\rm H}-3)}{(8-13X_{\rm H})}  
\sim \frac{(0.6E_{\rm g}-1.0)}{(2.7-E_{\rm g})}.  
\label{eq_REg}
\end{equation}
The above equation is an approximation to the exact relationship between $E_{\rm g}$, $X_{\rm H}$ and $R$, which depends upon the $sp^2$ aromatic domains sizes, the -CH$_3$ methyl group concentration and the particle size.\cite{2012A&A...540A...1J,2012A&A...540A...2J,2012A&A...542A..98J} The band gap is therefore a proxy for the a-C(:H) material optical properties of a given $R$ and $X_{\rm H}$.\cite{2012A&A...540A...1J,2012A&A...540A...2J}

It is therefore the evolution of the band gap that can be used to trace and characterise the inherent variability in the a-C(:H) optical properties and that is of prime importance in unravelling the evolution of hydrocarbon grains in the ISM.\cite{1990QJRAS..31..567J,1996MNRAS.283..343D,2009ASPC..414..473J, 2013A&A...558A..62J} Recent modelling work\cite{2012A&A...540A...1J,2012A&A...540A...2J,2012A&A...542A..98J} shows that the optical properties of a-C(:H) materials can be completely determined by two parameters, the band gap, $E_{\rm g}$, and the particle size, $a$, which turn out to be coupled for $a < 30$\,nm, in that the {\em minimum-possible} band gap is given by\cite{2012A&A...545C...3J}
\begin{equation}
E_{\rm g}(a)_{\rm min} = \left( \frac{1}{a \ {\rm [nm]}} - 0.2 \right) \ \ \ {\rm eV}, 
\label{eq_Ega}
\end{equation}
{\it i.e.}, the effective band gap, $E_{\rm g}({\rm eff})$, for a given size particle is 
\begin{equation}
E_{\rm g}({\rm eff}) = {\rm max} [ \, E_{\rm g}({\rm bulk}), \, E_{\rm g}(a)_{\rm min} \, ], 
\label{eq_Egeff}
\end{equation}
where $E_{\rm g}$(bulk) is the expected bulk a-C(:H) material band gap. For particles with radii larger than $\sim 30$\,nm this minimum band gap limit does not apply.

In this work we explore the likely consequences of interstellar a-C(:H) dust evolution in the transition between the diffuse ISM and the outer regions of molecular clouds ($n_{\rm H} = 10^2-10^4$\,cm$^{-3}$).

\subsection{Hydrocarbon dust photo-processing}

The (photo-)dissociation of CH bonds {\em within} a-C(:H) grains leads to the loss of H atoms from the aromatic, aliphatic and olefinic hydrocarbon structures in the chemical network, a progressive aromatisation and a closing of the optical band gap. This process is at the heart of the evolution of the physical and optical properties of a-C(:H) materials. Quantifying this fundamental process is therefore a key to advancing our understanding of the evolution of these materials in the ISM. Following earlier work\cite{2012A&A...540A...2J,2012A&A...545C...2J} the photo-darkening rate can be expressed as 
\begin{equation}
\Lambda_{\rm UV,pd}(a) = F_{\rm EUV} \ \sigma_{\rm CH} \ Q_{\rm abs}(a,E) \, \epsilon\ \ \ {\rm [\,s^{-1}]},  
\label{eq_UV_dehyd_rate}
\end{equation}
where $F_{\rm EUV} \simeq 3 \times 10^7$\,photons cm$^{-2}$ s$^{-1}$ is the CH bond-dissociating EUV photon flux in the local ISM \citep{2002ApJ...570..697H} (for the interstellar radiation field, ISRF, in the solar vicinity, {\it i.e.}, $G_0 =1$), $\sigma_{\rm CH} \simeq 10^{-19}$\,cm$^2$ is the CH bond photo-dissociation cross-section, \cite{1972JChPh..57..286W,1994CPL...227..243G,2001A&A...367..347M,2001A&A...367..355M} $Q_{\rm abs}(a,E)$ is the particle absorption efficiency factor in the optical cross-section determination, {\it i.e.}, $\sigma(a,E) = \pi a^2 Q_{\rm abs}(a,E)$ and $\epsilon$ is the photo-dissociation or photo-darkening efficiency. 

The a-C(:H) particle photo-processing time-scale, as a function of grain radius and the local ISRF, can then be expressed as 
\begin{equation}
\tau_{\rm UV,pd}(a,G_0) = \frac{1}{\Lambda_{\rm UV,pd}(a) \, G_0}. 
\label{eq_tau_UVpd}
\end{equation}

Fig.~\ref{fig_photoproc_time} shows the a-C(:H) particle photo-processing time-scale as a function of $\epsilon$ and grain radius.  As can be seen in this figure, the earlier estimates for $\tau_{\rm UV,pd}(a)$, with the assumption that $\epsilon = 0.1$ (diamonds),\cite{2012A&A...545C...2J} predict decreasing EUV processing time-scales with increasing particles size up to $a \simeq 30$\,nm, which are then approximately constant at around $(5 \times 10^4/G_0)$\,yr for $a > 30$\,nm. The increased photo-processing time-scales for $a < 30$\,nm are due to the decrease in the grain photon absorption efficiency with decreasing particle size, {\it i.e.}, $Q_{\rm abs} \propto a$ for particles smaller than the wavelength.

\begin{figure}[h]
\centering
 \includegraphics[height=8cm]{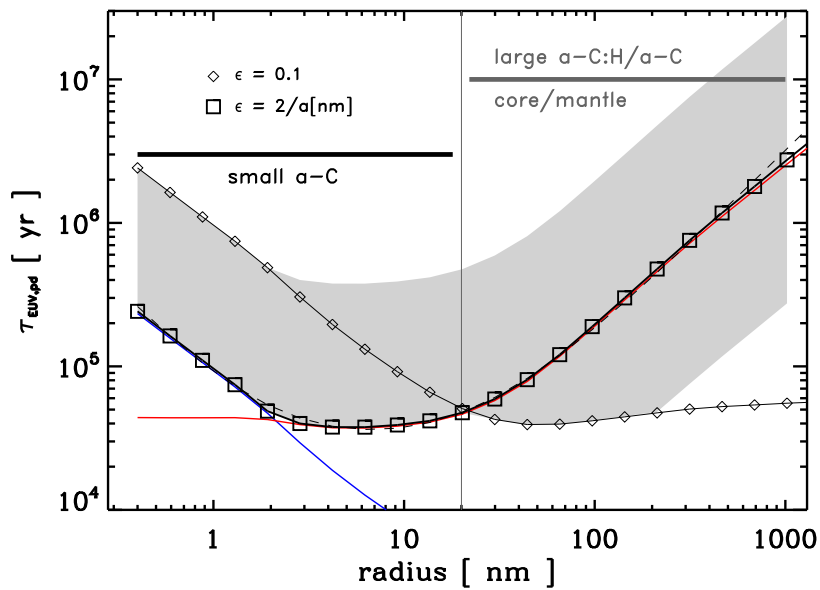}
 \caption{The a-C(:H) photo-processing time-scale for $G_0 = 1$ as a function of particle radius, $a$, and the CH bond photo-darkening efficiency, $\epsilon$ ($\epsilon = 0.1$, diamonds; $\epsilon = 2/a\,{\rm[nm]}$ for $a > 2$\,nm, Eq.(\ref{eq_epsa}), squares). 
 The vertical grey line indicates the critical radius between small homogeneous and larger core-mantle particles.  The grey band indicates an order of magnitude uncertainty on the derived time-scales, for grains with radii $\leq 200$\,nm the derived photo-processing time-scales are lower limits. 
 The blue and red lines indicate the small and large grain $(\epsilon \times Q_{\rm abs})^{-1}$ behaviours, {\it i.e.}, $\propto 1/Q_{\rm abs}$ and $\propto a/(2Q_{\rm abs}$), respectively. 
 The mostly-hidden, dashed grey line is the Eq.(\ref{eq_pd_timescale}) analytical fit to the derived a-C(:H) photo-processing time-scales (squares).}
 \label{fig_photoproc_time}
\end{figure}

We now introduce a size-dependent photo-dissociation or photo-darkening efficiency, $\epsilon(a)$, into our earlier derivation,\cite{2012A&A...540A...2J,2012A&A...545C...2J} and adopt the following  
\begin{equation}
\epsilon(a) = \left( \frac{2}{a\,{\rm[nm]}} \right) \ \ \ {\rm for} \ \ \ a > 2 \, {\rm nm} \ \ \ \ {\rm else} \ \ \ \ 
\epsilon(a) = 1 \ \ \ {\rm for}\ a \leq 2 \, {\rm nm}.  
\label{eq_epsa}
\end{equation} 
For a-C(:H) grains such a size-dependent CH bond photo-dissociation efficiency appears to be a physically more realistic scenario, than a fixed and size-in\-depend\-ent $\epsilon$. With increasing particle size other photon-driven processes will begin to compete with CH photo-dissociation and likely become more important in large grains ($a > 10$\,nm), {\it i.e.}, EUV photon absorption leading to thermal excitation and/or fluorescence. For small particles, $a \leq 2 \, {\rm nm}$, the CH bond photo-dissociation is determined by the absorption efficiency factor $Q_{\rm abs}$ (see Fig,~\ref{fig_photoproc_time}) and adopting $\epsilon(a) = 1$ therefore implies that every absorbed photon leads to CH bond breaking, as might be expected for particles in the `molecular domain'. If we now consider the effects of including the size-dependent photo-darkening efficiency, $\epsilon(a)$, given by Eq.~(\ref{eq_epsa}), the squares in Fig.~\ref{fig_photoproc_time}, we find a minimum in the processing time-scale, of the order of $(4 \times 10^4/G_0)$\,yr, for  particles with $a \sim 3-10$\,nm. For smaller and larger grains in the modelled dust size distribution the processing time-scale is $> (10^5/G_0)$\,yr. For particles smaller than $\simeq 3$\,nm the same trend as in the constant $\epsilon$ processing time-scales is apparent but it is shifted down by an order of magnitude because for $a < 2$\,nm we now assume $\epsilon(a) = 1$, rather than $\epsilon = 0.1$ as in the previous work. For particles with radii larger than $10$\,nm the processing time-scales increase with radius because of the assumed size-dependence of the photo-darkening efficiency $\epsilon(a) \propto a^{-1}$. 

The a-C(:H) grain photo-processing time-scales in the ISM, for a size-depend\-ent photo-dissociation efficiency, $\epsilon(a)$, can be analytically expressed as a function of the grain radius, $a$, in nm;  
\begin{equation}
\tau_{\rm UV,pd}(a) = \frac{10^4}{G_0} \bigg\{ 2.7  + \frac{6.5 }{(a\,[{\rm nm}])^{1.4}} + 0.04 \, (a\,[{\rm nm}])^{1.3}  \bigg\} \ \ {\rm [yr]}.    
\label{eq_pd_timescale}
\end{equation}
This fit to the predicted a-C(:H) photo-processing time-scales is shown by the somewhat hidden dashed line fit to the square data points in Fig.~\ref{fig_photoproc_time}.

\subsection{Hydrocarbon dust erosion}

In addition to EUV photo-processing in PDRs, which leads to the erosion of small carbonaceous particles,\cite{2012A&A...542A..69P} a-C(:H) dust will also be highly susceptible to processing, erosion and destruction in supernova(SN)-generated shock waves in the warm inter-cloud medium\cite{1994ApJ...433..797J,1996ApJ...469..740J,2008A&A...492..127S} Recent work\cite{2010A&A...510A..36M,2010A&A...510A..37M,2011A&A...526A..52M,2012A&A...545A.124B,2013A&A...556A...6B} shows that small a-C(:H) and poly-aromatic particles ({\it e.g}, polycyclic aromatic hydrocarbons, PAHs) are rather easily destroyed in SN shock waves, in a hot gas and by cosmic rays. This occurs, primarily, through the effects of electronic excitation and dissociation following electron and ion collisions, a process that is more important than direct ``knock-on'' sputtering in sub-nm- and nm-sized particles. The important conclusion of these works is that most of the small carbon population in energetic regions will be destroyed. The life-time of carbonaceous grains in the ISM that is derived from these studies is rather short, {\it i.e.}, $\leq 100$ million years. This catastrophic situation implies that carbonaceous dust must be efficiently re-formed in dense regions of the ISM where re-accretion re-forms carbonaceous matter principally in the form of mantles on the surviving silicate and any surviving carbonaceous grains.\cite{2011A&A...530A..44J}  The logical conclusion of this is that large pre-solar a-C(:H) grains should be rather rare. However, pre-solar carbon-rich grains of nano-diamond, SiC and graphite with anomalous isotopic compositions typical of of evolved stars and SNe are found in meteorites\cite{1993Metic..28..490A} but they are probably not the dominant carrier of the solid carbon phase.

\subsection{Hydrocarbon dust (re-)accretion}

Here we consider in detail the (re-)accretion of gas phase C and H atoms to form a-C(:H) carbonaceous mantles on dust in the transition between the diffuse and dense molecular ISM. In particular, we find that the nature of the mantles formed on dust in the transition to denser regions is critically-determined by the extinction at UV wavelengths and the hydrogenation of aromatic-rich carbonaceous grain materials.\cite{2002ApJ...569..531M,2006ApJ...647L..49M,2008ApJ...682L.101M,2010ApJ...718..867M} 

Here we explore the qualitative effects of mantle (re-)accretion in the outer regions of molecular clouds. We assume a semi-inifinite, planar cloud and adopt the simple density profile into the cloud from the surface,  
{\it e.g.}, similar to that derived for dense filaments \cite{2013A&A...559A.133Y} and photo-dissociation regions, \cite{2012A&A...541A..19A}
\begin{equation}
n_{\rm H}(l)    = n_{\rm H}(0) \, [ \, 1 + 10^5 l^2 \, ], 
\end{equation}
where $n_{\rm H}(0)$ is the cloud surface density in cm$^{-3}$, which we assume to be typical of the diffuse ISM, {\it i.e.}, $n_{\rm H}(0) = 40$\,cm$^{-3}$, and $l$ is the distance into the cloud from its surface in pc. 
The cloud density profile is shown in Fig. \ref{fig_cloud} along with the column density, $N_{\rm H}$, the optical depths $A_{\rm V}$ and $A_{\rm EUV}$ ($E_{\rm EUV} = 10$\,eV) and the attenuation of V and EUV band photons as a function of distance into the cloud. 
\begin{figure}[h]
\centering
  \includegraphics[height=7cm]{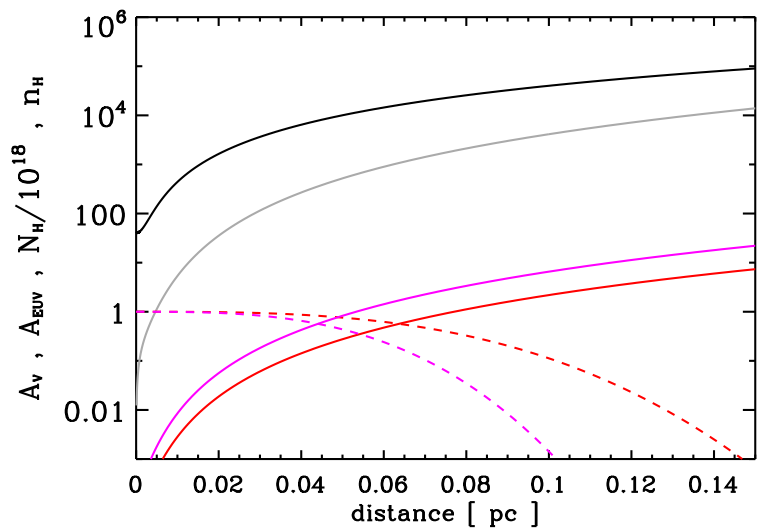}
  \caption{The cloud parameters: density $n_{\rm H}$ (H~cm$^{-3}$, black), column density $N_{\rm H}/10^{18}$ (H~cm$^{-2}$, grey), $A_{\rm V}$ (red), the attenuation at the V band wavelength $I/I_0 = e^{-A_{\rm V}}$ (dashed red), $A_{\rm EUV} = 3A_{\rm V}$ ($E_{\rm EUV} = 10$\,eV $\equiv 124$\,nm, violet, see Fig.~\ref{fig_ext}) and the attenuation at the EUV wavelength $I/I_0 = e^{-A_{\rm EUV}}$ (dashed violet),  as a function of distance into the cloud from its surface.}
  \label{fig_cloud}
\end{figure}

Recently, a completely new dust modelling approach was proposed, which does not include polycyclic aromatic hydrocarbons (PAHs), graphite grains or ``astronomical silicate''. \cite{2013A&A...558A..62J} Instead, it uses the size-dependent optical and thermal properties for hydrogenated amorphous carbons, a-C(:H),\cite{2012A&A...540A...1J,2012A&A...540A...2J,2012A&A...542A..98J,2012A&A...545C...2J,2012A&A...545C...3J} and the optical properties of an amorphous silicate with metallic iron inclusions.\cite{2013A&A...558A..62J} 
The new aspects in this model are: 
\begin{itemize}
  \item the continuum of a-C(:H) optical properties, 
  \item the FUV photo-processing of aliphatic-rich a-C:H into aromatic-rich a-C, 
  \item the a-C(:H) core-mantle structure of all carbon grains with radii $> 20$\,nm, 
  \item a-C mantled amorphous forsterite-type silicates, and
  \item the incorporation of iron as metal nano-inclusions into the silicate \\ (as a result of reduction by mantle carbon diffusion into the silicate\cite{2006A&A...448L...1D}).  
\end{itemize}
This new model satisfactorily explains the dust extinction, scattering and emission in the diffuse ISM and predicts the evolution of the carbonaceous dust properties in response to the local conditions.\cite{2013A&A...558A..62J} 

In this work we apply the Jones {\it et al.} dust model,\cite{2013A&A...558A..62J} using the extinction and mass size distributions for the model derived using the DustEM tool,\cite{2011A&A...525A.103C}  as shown in Figs. \ref{fig_sdist} and \ref{fig_ext}. For this model the gas phase C atom accretion timescale to form a-C(:H) mantles is given by
\begin{equation}
t_{\rm acc}  = [ \, \Sigma_{\rm total} \ n_{\rm H} \ X_{\rm C} \ v_{\rm C} \ S_{\rm C} \, ]^{-1}, 
\end{equation}
where $\Sigma_{\rm total}$ is the total dust cross-section per H atom, $X_{\rm C}$ is the relative abundance of gas phase carbon atoms and $v_{\rm C} =  [8k_{\rm B}T_{\rm kin}/(\pi m_{\rm C})]^{0.5}$ is their thermal velocity ($k_{\rm B}$ is the Boltzmann constant, $T_{\rm kin} = 80$\,K is the gas kinetic temperature and $m_{\rm C}$ the C atom mass), and $S_{\rm C}$ is the C atom sticking coefficient (assumed to be unity).  We assume a cosmic abundance of $\sim 400$\,ppm for carbon,\cite{2012ApJ...760...36P} the dust model requires 233\,ppm\cite{2013A&A...558A..62J} therefore leaving $\sim 167$\,ppm available for accretion onto dust in the form of a-C(:H) mantles.  

It has been shown that H atom incorporation into carbonaceous materials at low temperature ($T_{\rm kin} = 80$\,K) can lead to their hydrogenation\cite{2002ApJ...569..531M,2006ApJ...647L..49M,2008ApJ...682L.101M,2010ApJ...718..867M} and we therefore consider this possibility within the framework of our qualitative model. The H atom sticking time-scale onto all grains is 
\begin{equation}
t_{\rm H}  = [ \, \Sigma_{\rm total} \ n_{\rm H} \ v_{\rm H} \ S_{\rm H} \, ]^{-1}, 
\end{equation}
where $v_{\rm H} =  [8k_{\rm B}T_{\rm kin}/(\pi m_{\rm H})]^{0.5}$ is the H atom thermal velocity and $S_{\rm H}$ is the H atom sticking coefficient, here we assume the canonical value $S_{\rm H} = 0.3$. We define the low temperature a-C(:H) grain (re-)hydrogenation rate as 
\begin{equation}
t_{\rm Hin}  = [ \, \xi \ \Sigma_{\rm total} \ n_{\rm H} \ v_{\rm H} \ S_{\rm H} \, ]^{-1}, 
\end{equation}
where $\xi$ is the efficiency for H atom incorporation into the a-C(:H) structure. 
Given that most incident H atoms will combine with other incident H atoms to form H$_2$, which is then ejected, $\xi$ should be significantly less than unity. 

For the size distribution in the Jones {\it et al.} standard diffuse ISM dust model\cite{2013A&A...558A..62J} we derive $\Sigma_{\rm total} = 7.3\times 10^{-21}$\,cm$^{2}$/H atom and find that the accretion time-scale and mantle thickness are very dependent upon the lower grain size limit. For example, the accretion of $\sim 170$\,ppm of carbon onto all grain surfaces in the standard diffuse ISM dust model, yields $t_{\rm acc} \sim 10^7$\,yr for $n_{\rm H} = 10^4$\,cm$^{-3}$ and a mantle thickness of $\simeq 3$\,nm on all grains. However, in the outer regions of molecular clouds the small grains ($a \leq 5$\,nm) are accreted/coagulated onto large grains.\cite{2003A&A...398..551S,2012A&A...548A..61K,2013A&A...559A.133Y} As an illustration of the important effects of small particles, if we remove all grains with radii $<5$\,nm from the dust size distribution we find a carbonaceous mantle thickness of $160$\,nm, a reduction in the total grain cross-section by more than an order of magnitude, to $\Sigma_{\rm total} = 1.4\times 10^{-22}$\,cm$^{2}$/H atom, and a corresponding increase in the C atom accretion time-scale to $t_{\rm acc} \geq 10^9$\,yr for $A_{\rm V} < 3$. Thus, if carbon mantle accretion onto nano-particles is inhibited by stochastic heating events or if the nano-particles themselves are coagulated/accreted onto or into large particles, the accreted mantle thickness could be $\gg 3$\,nm but the accretion time-scales would then be $> 10^7$\,yr. 
\begin{figure}[t]
\centering
  \includegraphics[height=7cm]{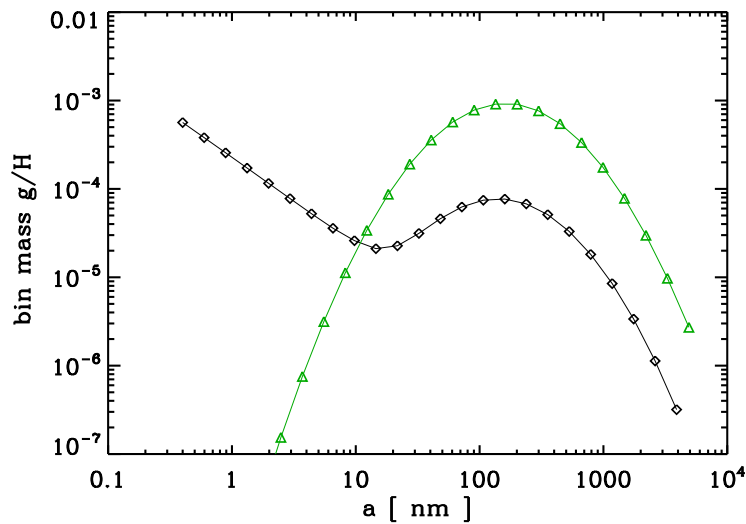}
  \caption{The model dust mass/size distribution in the diffuse ISM: a-C(:H) grains (black) and a-Sil$_{\rm Fe}$/a-C core/mantle grains (green).}
  \label{fig_sdist}
\end{figure}
\begin{figure}[t]
\centering
  \includegraphics[height=7cm]{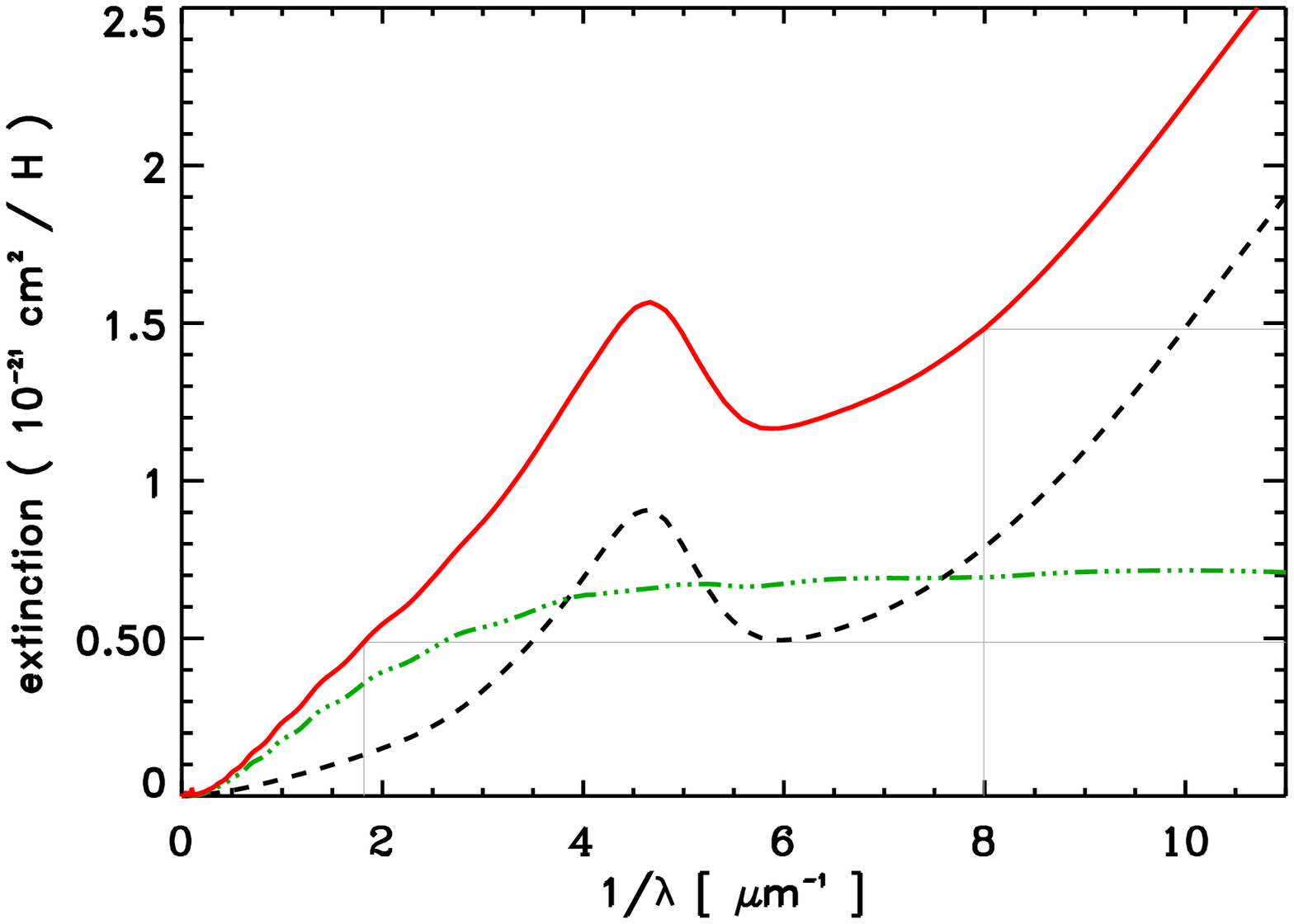}
  \caption{The modelled dust extinction in the diffuse ISM: a-C(:H) grains (black) and a-Sil$_{\rm Fe}$/a-C core/mantle grains (green). The total extinction is given by the red line. The grey lines indicate the extinction in the V band and for 10\,eV EUV photons.}
  \label{fig_ext}
\end{figure}

Interestingly, the analysis of the $\sim 100$\,nm thick ``organic'' coatings on the mineral grains in interplanetary dust particles (IDPs) indicates that this primitive matter is the result of the condensation of C-bearing ``ices'' onto grain surfaces and the formation of refractory matter by subsequent UV or other ionising irradiation.\cite{2010cosp...38.3242F} These ``organic'' mantles, which carry H and N isotopic anomalies consistent with molecular cloud or outer Solar System material, resemble the a-C(:H) mantles on the grains in the Jones {\it et al.} diffuse ISM dust model\cite{2013A&A...558A..62J} and could be the result of the direct accretion of the remaining gas phase carbon as a-C:H, as discussed here, rather than requiring a carbon-rich ice precursor phase. Such ``organic'' coatings on carbonaceous grains would lead to large aliphatic-rich a-C:H grains that resemble the ``organic globules'' detected in primitive meteorites and comets, which show D/H and $^{15}$N/$^{14}$N enrichments, and that are likely the product of low temperature ($\sim 10$\,K) chemical reactions in cold molecular clouds or the outer regions of the proto-solar nebula.\cite{2008cosp...37.2018M}

In Fig.~\ref{fig_tproc} we show the time-scale for C atom accretion into a-C(:H) on all grains, $t_{\rm acc}$, and the a-C(:H) grain EUV photo-processing time-scale, $t_{\rm pd}$, for 1, 10 and 100\,nm radius particles, as a function of distance into the cloud and $A_{\rm V}$. We deduce from Fig.~\ref{fig_tproc} that the accretion and EUV photo-processing time-scales are similar for cloud depths of the order of $\sim 0.09-0.1$\,pc, equivalent to $A_{\rm V} \sim 2$. 
Nearer to the cloud surface, $A_{\rm V} < 2$ where $t_{\rm acc} \gg t_{\rm pd}$, it is evident that grain mantling can only occur on time-scales $\geq 10^7$\,yr. In the absence of any incident H atom hydrogenation of the grains, the outer grain surfaces/mantles will be EUV-photolysed to H-poor, aromatic-rich materials, {\it i.e.}, low band gap a-C with $E_{\rm g} < 0.2$\,eV. However, deeper into the cloud the CH bond photo-dissociating EUV photons are attenuated and the cloud interiors, $A_{\rm V} > 2$, are UV-poor regions (UVPRs) where  a-C:H grains and mantles will not be EUV photo-processed because the photolysis time-scale is significantly longer than the C (and H) atom accretion timescale. Thus, in UVPRs the accreted mantles will be H-rich, aliphatic-rich materials, {\it i.e.}, wide band gap a-C:H with $E_{\rm g} \geq 2$\,eV. 
\begin{figure}[h]
\centering
  \includegraphics[height=7cm]{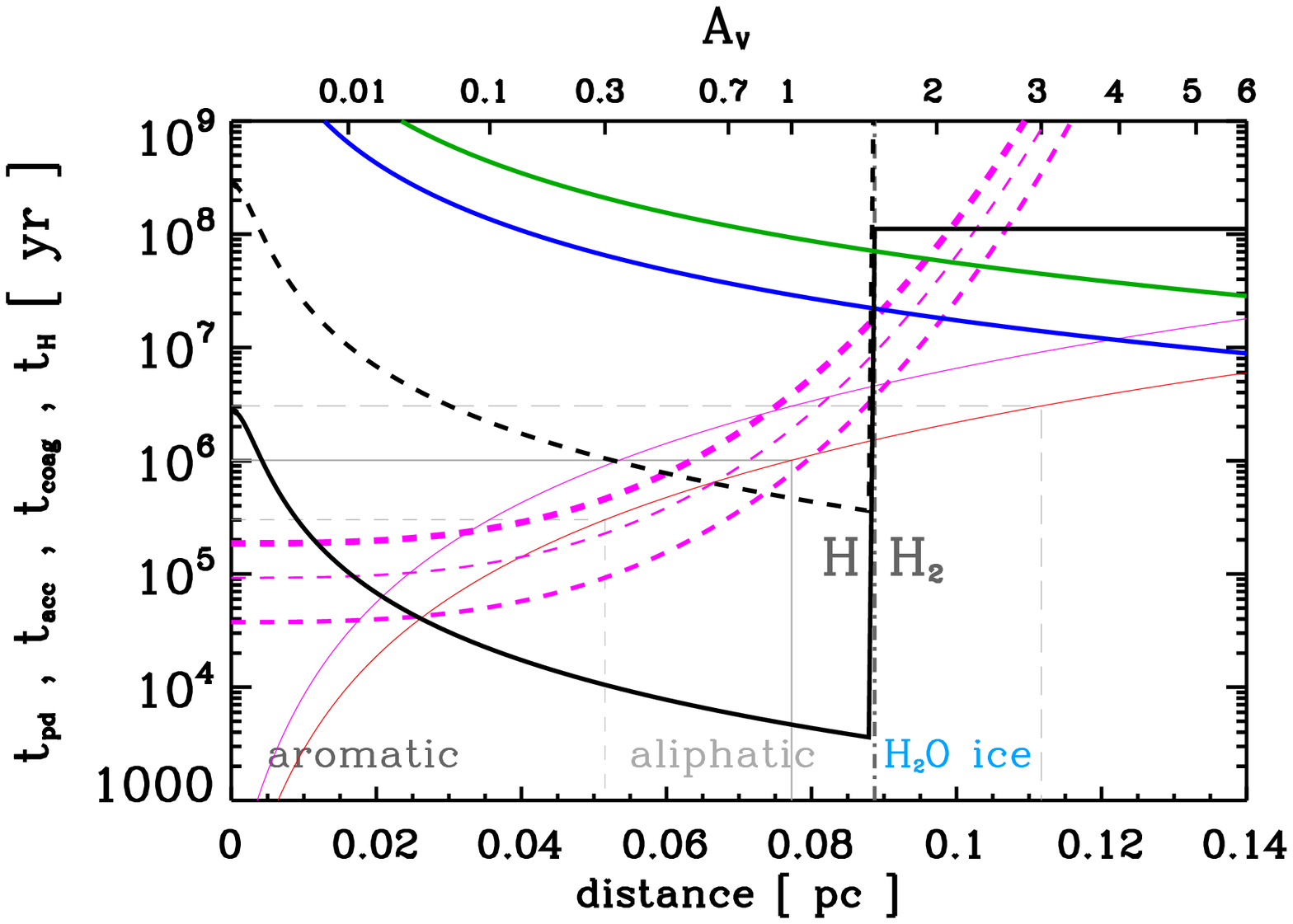}
  \caption{The dust processing time-scales as a function of $A_{\rm V}$ and distance into the cloud: the C atom accretion time-scale $t_{\rm acc}$ (blue), the small a-C grain-large silicate grain coagul\-ation time-scale $t_{\rm coag}$ (green), the H atom sticking timescale, $t_{\rm H}$ (black), the H atom incorp\-oration timescale, $t_{\rm Hin}$ (black dashed), and the EUV photo-processing time-scale for 1, 10 and 100\,nm radius a-C(:H) grains $t_{\rm pd}$ (thin, medium and thick dashed violet lines respectively). Also shown are $A_{\rm V}$ (red) and $A_{\rm EUV} = 3A_{\rm V}$ (violet) multiplied by $10^6$. The grey lines indicate the depth into the cloud at which $A_{\rm V} \sim 0.3, 1$ and 3, dashed, solid and dashed, respectively. The vertical dash-dotted line indicates the atomic to molecular hydrogen transition (here assumed to occur at $A_{\rm V} = 1.5$). The approximate aromatic, aliphatic and H$_2$O ice accreted mantle regimes are also indicated.}
  \label{fig_tproc}
\end{figure}

We now consider the possibility that H atom hydrogenation is efficient at low temperatures.\cite{2002ApJ...569..531M,2006ApJ...647L..49M,2008ApJ...682L.101M,2010ApJ...718..867M} Fig.~\ref{fig_tproc} indicates that in the low density outer regions of clouds, with $A_{\rm V} < 0.01$, the H atom collision rate is insufficient to counteract the EUV photo-dissociation of the CH bonds within a-C(:H), even if every H atom that sticks to the grains is incorporated into the a-C(:H) structure, {\it i.e.}, $\xi = 1$. Adopting a significantly lower efficiency, $\xi = 0.01$, leads to the same result but the depth for the onset of hydrogenation shifts deeper into the cloud ($A_{\rm V} \geq 0.7$). These results imply that in the low extinction diffuse ISM and outer cloud regions ($A_{\rm V} < 0.01$ for $\xi = 1$ or $A_{\rm V} < 0.7$ for $\xi = 0.01$, with $n_{\rm H} \sim 40$\,cm$^{-3}$ and $T_{\rm kin} \sim 80$\,K) small a-C(:H) particles and any a-C(:H) on the surfaces/mantles of large grains must be of predominantly H-poor, aromatic-rich, a-C materials. Deep within the clouds hydrogen is predominantly in molecular form for $A_{\rm V} \geq 1.5$,\cite{1985ApJ...291..722T} except for a background H atom abundance of $\sim 1$\,cm$^{-3}$  due to the effect of cosmic rays, and so the hydrogenation process will be reduced by orders of magnitude in a molecular gas. However, as Fig.~\ref{fig_tproc} shows, at intermediate optical depths, $0.01-0.7 \leq A_{\rm V} \leq 1.5$ (the lower limit depends on the assumed value for $\xi$), $t_{\rm Hin} < t_{\rm pd} < t_{\rm acc}$, {\it i.e.}, the H atom incorporation into a-C(:H) is faster than both CH bond photo-dissociation and mantle accretion. Thus, the transformation of aromatic-rich grain materials into H-rich, aliphatic-rich materials could occur on time-scales of the order of $10^6$\,yr. Somewhat deeper into the cloud, $\sim 1.5 \leq A_{\rm V} \leq 2.5$, and immediately after the transition from atomic to molecular hydrogen, H atom incorporation is again subservient to photo-dissociation (albeit at a very low level) 
and any remaining gas phase carbon in the clouds would accrete
along with the ice mantles in regions with $A_{\rm V} \geq 1.5$. 

\subsection{Grain coagulation effects}

Following earlier work\cite{2003A&A...398..551S} the coagulation time-scale between dust species 1 and 2 can be expressed as 
\begin{equation}
t_{\rm coag}  = \left\{ \pi(a_1+a_2)^2 \ X_1 \ n_{\rm H} \ \Delta v \right\}^{-1}, 
\end{equation}
where $a_i$ is the grain radius, $X_1$ is the relative abundance of dust species 1 and $\Delta v$ is the relative velocity between the particles. We are interested in the time-scale for large grains to sweep up small aromatic-rich grains, in an accretion-type coagulation process\cite{2003A&A...398..551S,2011A&A...528A..96K,2012A&A...548A..61K,2013A&A...559A.133Y} that leads to the formation of aromatic particle mantles. In order to estimate the coagulation time-scale we set $a_1 = 5$\,nm and $a_2 = 160$\,nm, similar to previous estimates\cite{2012A&A...548A..61K} and based on the dust masses in the new model\cite{2013A&A...558A..62J}. The small-large grain coagulation time-scale is shown in Fig. \ref{fig_tproc} and indicates that small grain coagulation always lags behind C atom accretion as a viable a-C(:H) mantling process in cloud interiors.

\begin{figure}[h]
\centering
  \includegraphics[height=11cm,angle=270]{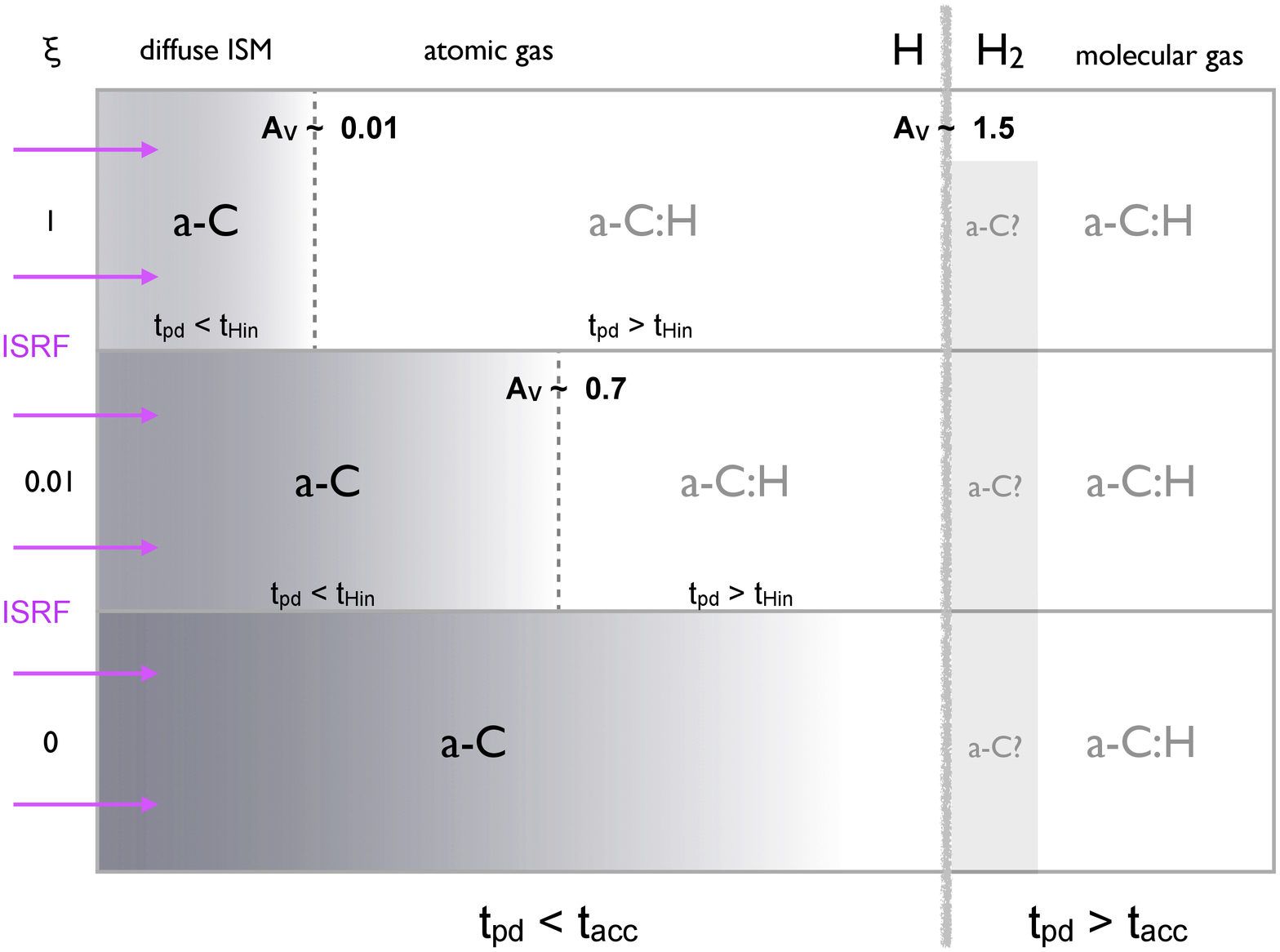}
  \caption{A schematic view of the accreted/transformed carbonaceous grain/mantle composition as a function of the optical depth, $A_{\rm V}$, and the H atom incorporation efficiency into a-C(:H), $\xi$.}
  \label{fig_schema}
\end{figure}
\section{Astrophysical implications}

Fig.~\ref{fig_schema} gives a schematic view of  the evolution of the accreted/transformed a-C(:H) grain/mantle composition as a function of the optical depth, $A_{\rm V}$, and the H atom incorporation efficiency into a-C(:H), $\xi$. This figure indicates, and Fig.~\ref{fig_tproc} shows, that in the diffuse ISM and in the outer regions of molecular clouds, the time-scale for H atom incorporation into a-C(:H) grains is significantly longer than the CH photo-dissociation time-scale. In low extinction regions ($A_{\rm V} < 0.01-0.7$) H atom incorporation into a-C(:H) is therefore seemingly not fast enough to transform a-C into a-C:H. Thus, given that the EUV CH photo-diss\-ociation depth is of the order of 20\,nm,\cite{2012A&A...540A...2J} this work indicates that small a-C(:H) grains and the outer surfaces of large a-C(:H) grains will be rapidly transformed ($t_{\rm pd} \geq 4 \times 10^4$\,yr) to aromatic-rich a-C (see Fig.~\ref{fig_tproc}),  as in the recently-proposed dust model.\cite{2013A&A...558A..62J} 
The cores of  large a-C(:H) particles ($a \gg 20$\,nm) being shielded from the effects of CH bond EUV photo-dissociation can therefore be of H-rich, aliphatic-rich a-C:H material.\cite{2012A&A...540A...2J} Clearly, the cores of these large a-C(:H) grains can only be H-rich in the diffuse ISM if they were formed as such because they cannot be re-hydrogenated there. However, in slightly higher extinction regions ($A_{\rm V} > 0.01-0.7$) any grain core a-C material could be transformed into a-C:H by the effects of H atom incorporation\cite{2002ApJ...569..531M,2006ApJ...647L..49M,2008ApJ...682L.101M,2010ApJ...718..867M} if this process is efficient and if they are not already H-rich. Interestingly, as shown in Fig.~\ref{fig_tproc},  somewhat deeper into a cloud than the atomic to molecular hydrogen interface, at $A_{\rm V} \sim 2$, the time-scale for H atom incorporation into a-C increases above that for CH bond photo-dissociation and so any accreting a-C(:H) mantles would be aromatic-rich rather than aliphatic-rich. This would seemingly occur 
with the onset of ice mantle formation for $A_{\rm V} \geq 1.5$.

It has been proposed that in the ISM the carbonaceous grains consist of aliphatic-rich mantle overlying an aromatic-rich core.\cite{2013ApJ...770...78C} This is the inverse of the Jones {\it et al.} dust model\cite{2013A&A...558A..62J} and is at odds with our results for dust processing in the diffuse ISM. Nevertheless, within  molecular clouds it is indeed likely that the carbonaceous grains surfaces will be aliphatic-rich due to the effects of a-C:H mantle accretion and/or the incident H atom aliphatisation of aromatics. 

The formation of H-rich, aliphatic-rich mantles/grain surfaces, through a-C:H mantle accretion or H atom incorporation into a-C grains, leads to a slight decrease in the visible extinction, a diminution of the UV bump but little change in the FUV extinction.\cite{2013A&A...558A..62J}  This is consistent with the high carbon depletion ($\sim 400$\,ppm of C in dust) along the line of sight towards HD\,207198\cite{2012ApJ...760...36P, 2013A&A...558A..62J} but inconsistent with the extinction curve along lines of sight through the denser diffuse ISM, with $R_{\rm V} \sim 5$, where the UV extinction is flatter than the average, {\it i.e.}, for lines of sight with $R_{\rm V} \sim 3$. The flatter UV extinction along high $R_{\rm V}$ lines of sight implies that the coagulation of small grains onto large grains occurs before or is contemporaneous with carbonaceous mantle accretion. 
However, this preliminary study seems to indicate that accretion occurs before small grains coagulate onto large grains, {\it i.e.}, $t_{\rm coag}$ is always $> t_{\rm acc}$. The extinction curve variations and far-IR to sub-mm dust emission generally seem to imply that small grains stick onto big grains before significant accretion occurs.\cite{2012A&A...548A..61K} An exception to this seems to be the HD\,207198 line of sight\cite{2012ApJ...760...36P, 2013A&A...558A..62J} where the formation of a-C:H mantles, or the transformation of a-C to a-C:H by H atom incorporation,\cite{2010ApJ...718..867M} is strongly implied. Detailed studies of the evolution of dust evolution in the low density ISM, through the effects of coagulation and accretion, will be required in order to unravel the exact sequence of events as a function of the local conditions.

\section{Summary and conclusions}

Carbonaceous dust in the ISM is apparently a much more fragile dust component than previously thought and must therefore be efficiently re-formed in the ISM. In this work we have investigated the composition and evolution of a-C(:H) dust in the transition from the low-density diffuse ISM to the outer reaches of molecular clouds. The primary processes that drive this evolution are the CH bond EUV photo-dissociative aromatisation, a-C:H $\rightarrow$ a-C, and the reverse process of H atom incorporation and aliphatisation, a-C $\rightarrow$ a-C:H, leading to (re-)hydrogenation.  We find that in the low-density diffuse ISM ($A_{\rm V} < 0.01-0.7$) photo-processing dominates over aliphatisation and that small carbonaceous grains/thin mantles and the outer surfaces of large carbonaceous grains should be predominantly of low band gap ($E_{\rm g} \sim 0$\,eV) aromatic-rich a-C materials. However, in the diffuse ISM, where H atom (re-)hydrogenation is slower than EUV photo-aromatisation, the carbonaceous grain cores can be of H-rich aliphatic-rich material if they were formed as such. In slightly higher extinction regions ($A_{\rm V} > 0.01-0.7$) we find that H atom incorporation into a-C leading to aliphatisation and an opening of the band gap could dominate. Interestingly, somewhat deeper into a cloud than the H/H$_2$ interface ($A_{\rm V} \sim 2$) any accreting gas phase carbon will tend to form a-C rather than a-C:H mantles because of the low H atom abundance. This unusual effect seemingly occurs 
at about the same optical depth as the accretion of ice mantles (at $A_{\rm V} \geq 1.5$; assuming a semi-infinite slab cloud model), which merits a more detailed analysis than has been possible here. 

In order to fully quantify the effects studied here a much more detailed study of the equilibrium composition of a-C(:H) materials in the ISM is required. This will necessitate a much deeper understanding of the spatio-temporal evolution of the dust composition ({\it e.g.}, band gap, $E_{\rm g}$) and the dust size distribution resulting from the effects of CH bond photo-dissociation and a-C(:H) grain photo-fragmentation in the low-density ISM and the effects of accretion and coagulation in denser regions. In particular, a much deeper understanding of the interplay between accretion and coagulation is required. Allied to this must be a study of the likely effects of turbulence and radiation pressure on the accretion onto dust and grain coagulation in the outer regions of molecular clouds. Also, given that C$^+$ remains the most abundant form of gas phase carbon to optical depths $A_{\rm V} \sim 3$\cite{1985ApJ...291..722T} and that the grains are negatively charged grains in molecular clouds the collision cross-sections for accretion could be somewhat enhanced within clouds. 
However, as is often assumed, ion accretion onto negatively-charged grains leads to recombination and the desorption of the incident ion. Thus, any accretion gain could be offset by C$^+$ recombination upon collision.
These processes could play an important role in dust evolution in these transition regions because they would tend to enhance the gas-grain and grain-grain collision velocities and therefore reduce the relevant time-scales. However, it should be noted that small grain coagulation onto large grains leads to a reduced total grain surface and therefore thicker mantles but also to significantly longer accretion time-scales.

The carbonaceous mantle accretion phenomena investigated here ought to have important observable consequences. Thus, the outer regions of molecular clouds ($\sim 0.5 \leq A_{\rm V} \leq 3$) merit deeper investigation because they will provide strong constraints on the nature and evolution of the matter that accretes onto dust and on the coagulation process. 

\section*{Acknowledgements}
This research was, in part, made possible through the financial support of the Agence National de la Recherche (ANR) through the programme CIMMES (ANR-11-BS56-029-02).

\footnotesize{
\bibliography{biblio_HAC} 
\bibliographystyle{rsc}
}

\end{document}